\documentclass[journal]{IEEEtran}

\usepackage{cite}
\usepackage{amsmath,amssymb,bm}
\usepackage{graphicx}
\usepackage{url}
\usepackage{siunitx}
\usepackage{caption}
\usepackage{subcaption}
\usepackage{color}
\usepackage{float}

\title{Time Domain Design of a Josephson Parametric Amplifier and Comparison with Input--Output Theory}

\author{%
  Emre~Küçükyılmaz, Mehmet~Ünlü, and Ali~Bozbey\\
  Department of Electrical and Electronics Engineering, TOBB University of Economics and Technology, Ankara, Türkiye\\
  Email: emrekucukyilmaz@etu.edu.tr
}

\begin{document}
\maketitle

\begin{abstract}
Quantum-limited amplifiers, such as Josephson Traveling Wave Parametric Amplifiers (JTWPAs) and Josephson Parametric Amplifiers (JPAs), are essential components in quantum computers. They amplify low-power microwave signals from qubits at the 10 mK stage before further amplification at the 4 K stage using HEMT amplifiers. In JPAs, parametric amplification is based on the nonlinear properties of Josephson Junctions. While JPAs are typically designed and analyzed using input-output theory based on quantum physics, we propose an alternative approach based on an equivalent circuit model of JPAs, implemented using open-source Josephson circuit simulators. We compare the results with those obtained from input-output theory. This method enables the use of circuit optimizers for various objective functions and significantly reduces design time compared to quantum theory-based approaches.
\end{abstract}


\section{Introduction}

In the literature, Josephson parametric amplifiers have been extensively investigated \cite{Castellanos_Mak,Planat_Sat_PWR,Ranzani,grebel,elo}. They are used as quantum limited amplifiers or as a non-classical light source and have multiple modes of operation such as degenerate, non-degenerate, or 4 wave mixing and 3 wave mixing modes \cite{Widely_Tun_Nondeg_3WM_SQL,Circuits_for_JPA_Qradar,Engineered_JPA_2M_SQZ}. Although they are characterized as weakly nonlinear quantum harmonic oscillators rather than strongly nonlinear qubits, they essentially resemble the same architecture \cite{Higher_Order_Effect}. Input-output theory, a widely used framework in quantum optics, models the dissipative interaction of the cavity with its surrounding environment, which is treated as a set of independent harmonic oscillators \cite{IO_Theory}. This approach operates in the Heisenberg picture, enabling the derivation of equations of motion for the intra-cavity field. These can then be related to the input and output fields via boundary conditions. The formalism remains accurate even in the presence of coupled or complex cavity configurations \cite{Eichler,Graph_based_analysis}. Interestingly, the semi-classical approach used in the analysis naturally lends itself to implementation with circuit simulators\cite{JPA_HBM}. In this work, we use open source Josephson circuit simulators such as Jsim and JoSIM \cite{JOSIM,Coldflux} to characterize the device and compare the results with linearized and generalized quantum treatment.We also derive the equation of motion for JPA flux and compute it in Matlab to demonstrate the agreement between two different numerical method-based solvers.In the literature, Josephson circuit simulators are widely recognized within the digital design community \cite{askerzadeJJ,JJ_SOMA,tukel_development_2013,tukel_optimization_2013,Coenrad_Experimental_Verif_of_Moat,Semenov, nakamura2009current}. However, RF behavior can also be analyzed using circuit simulators, as RF circuits can be accurately represented with lumped element models when modeled correctly. Motivated by this, we anticipate that quantum engineers will integrate time-domain Josephson simulators into their design workflow, simplifying and improving their methodology while advancing the development of scalable quantum computers.

\section{JPA Design Based on Input--Output Theory}
\label{IOT}

\begin{figure}[H]
\centering
\includegraphics[width=0.45\textwidth, trim=0.35cm 0.35cm 0.35cm 0.35cm, clip] {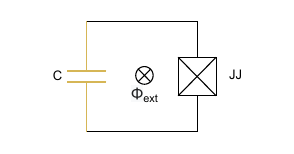}
\caption{\label{fig:12} Circuit considered for quantization}  
\end{figure}

\noindent To derive a Hamiltonian of an arbitrary circuit, we follow the steps explained in \cite{int_to_qec}. We represent the JPA as a parallel non-linear LC circuit and use the highlighted capacitive spanning tree to derive the Hamiltonian of the circuit given in figure \ref{fig:12} to derive the associated Hamiltonian as in equation (\ref{eq:DCSQ Hamiltonian}).

\begin{equation}
\mathcal{H} = \frac{C}{2} \dot{\Phi}^2 - E_{J} \cos\left(\frac{2\pi}{\Phi_0} (\Phi - \Phi_\text{ext})\right) 
\label{eq:DCSQ Hamiltonian}
\end{equation}
Where, $\Phi$ is the flux, $\Phi_0$ is the single flux quantum, $\Phi_{ext}$ is the external flux, $E_J$ is the Josephson energy. In the equation, $1^{st}$ term represents the energy associated with the capacitor, $C$, and $2^{nd}$ term represents the energy associated with the nonlinear inductor implemented by a Josepshon junction.   And because single junction potential is symmetric, RF driving the circuit is enough to induce the 4 wave mixing process without triggering any 3 wave mixing process which makes it convenient to drop the external flux dependence. In this case, one obtains the Hamiltonian given in equation (\ref{eq:DCSQ Simplified hamiltonian}).

\begin{equation}
\mathcal{H} = \frac{C}{2} \dot{\hat{\Phi}}^2 - E_J \cos \left( \frac{2 \pi}{\Phi_0} \hat{\Phi} \right)
\label{eq:DCSQ Simplified hamiltonian}
\end{equation}
This Hamiltonian can be quantized using Dirac's method. First, we introduce flux ($\hat{\Phi}$) and charge operators ($\hat{Q}$):

\begin{equation}
\begin{aligned}
\text{(a)} \quad & \hat{\Phi} = \Phi_\text{zpf} (\hat{a}^\dagger + \hat{a}) = \sqrt{\frac{\hbar Z_0}{2}} (\hat{a}^\dagger + \hat{a}), \\
\text{(b)} \quad & \hat{Q} = i Q_\text{zpf} (\hat{a}^\dagger - \hat{a}) = i \sqrt{\frac{\hbar}{2Z_0}} (\hat{a}^\dagger - \hat{a}).
\end{aligned}
\label{eq:annihilation,creation operators}
\end{equation}

\noindent Where $\hat{\Phi}_{zpf}$ and $\hat{Q}_{zpf}$ represent the zero-point fluctuations of the flux and change. The ladder operators \( \hat{a} \) and \( \hat{a}^\dagger \) satisfy the usual bosonic commutation relation,
\( [\hat{a}, \hat{a}^\dagger] = 1 \) and the canonical conjugate variables \( \hat{\Phi} \) and \( \hat{Q} \) satisfy the commutation relation \([ \hat{\Phi}, \hat{Q} ] = i \hbar\).
If we expand the cosine term in equation (\ref{eq:DCSQ Simplified hamiltonian}) and apply a rotating wave approximation assuming that the rotating terms do not have strong influence on the system, we obtain:

\begin{equation}
\begin{aligned}
\mathcal{H} =\frac{\hat{Q}^2}{2 C_{\text{}}} + \frac{\hat{\Phi}^2}{2 L_J} - \frac{E'_J}{4!} \left( \frac{2 \pi \hat{\Phi}}{\Phi_0} \right)^4 + O\left( \left( \frac{2 \pi \hat{\Phi}}{\Phi_0} \right)^6 \right)
\end{aligned}
\label{Hamiltonian, Charge/Flux Basis}
\end{equation}

\noindent Where $L_J$ is the Josephson inductance. Substituting $\hat{\Phi}$ and $\hat{Q}$  in equation (\ref{Hamiltonian, Charge/Flux Basis}) we obtain the so-called second quantization form as follows:

\begin{equation}
\begin{aligned}
\mathcal{H}^{\text{RWA}} = \hbar \tilde{\omega}_0 \hat{a}^\dagger \hat{a} + \frac{K}{2} \hat{a}^\dagger \hat{a}^\dagger \hat{a} \hat{a}
\end{aligned}
\label{Sec quant. Hamiltonian}
\end{equation}

\noindent Where \( \tilde{\omega}_0 = \omega_0 + K \) is the shifted frequency arising from zero-point fluctuations (ZPF). From the second quantized Hamiltonian obtained at equation \ref{Sec quant. Hamiltonian}, one can use the Heisenberg equation of motion (HEOM) to derive the quantum Langevin equation (\ref{IO theory}),  \cite{Quintana}.

\begin{equation}
\begin{aligned}
\dot{A} = -i \tilde{\omega}_0 A - i K A^\dagger A A - (\gamma_1 + \gamma_2) A \\ + \sqrt{2\gamma_1} a_{\text{in}}(t) + \sqrt{2\gamma_2} b_{\text{in}}(t)
\end{aligned}
\label{IO theory}
\end{equation}

\noindent Here, the operators \( A \), \( a_{\text{in}} \), and \( b_{\text{in}} \) represent the resonator, the input signal, and the input noise modes, respectively. Here, the signal and noise linewidths ($\gamma_1$ and $\gamma_2$ respectively) are defined as half width at half maximum. Considering only the pump input and neglecting any signal input for equation (\ref{IO theory}), In steady state i.e ($\dot{A}$=0) we obtain equation (\ref{duffingresponse}). 

\begin{equation}
\begin{aligned}
K^2 N^3 &+ 2(\omega_0 - \omega_p)K N^2 \\
&+ \left[(\omega_0 - \omega_p)^2 + (\gamma_1 + \gamma_2)^2\right] N
= 2 \gamma_1 p_{\mathrm{in}}^2
\end{aligned}
\label{duffingresponse}
\end{equation}

\noindent As this equation accounts solely for the presence of the pump wave,
N represents the number of pump photons confined within the cavity. This parameter is crucial, as it is the pump wave that initiates and sustains the parametric amplification process. Symbol $p_{in}$ is pump power. Since this equation is a cubic equation in N, it has multiple real solutions for certain drive powers. Some solutions may be in the bistable regime, which is known as the bifurcation regime, and has been treated in the literature\cite{Vijay_Review,Vijay_Siddiqi}.

\begin{figure}[H]
\centering
\includegraphics[width=\linewidth,
trim=0cm 0cm 0cm 1.5cm, clip]{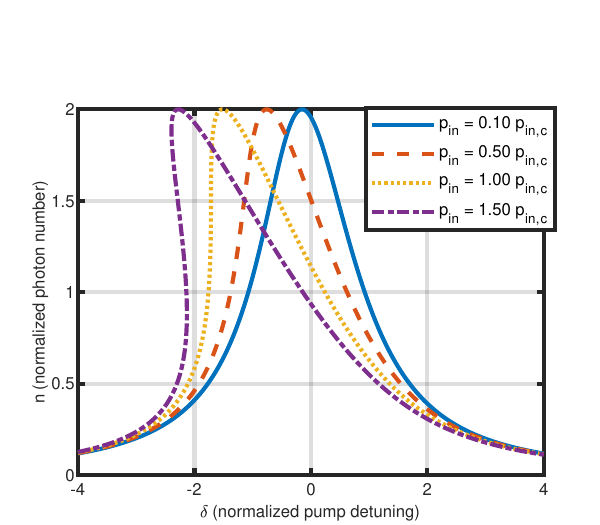}
\caption{\label{fig:1}  Normalized photon number (n) vs. normalized pump detuning ($\delta$) for various drive powers. $n$ = N / $(p_{in})^2 \frac{\gamma_1}{(\gamma_1 + \gamma_2)^2}$  and $\delta$ = $\omega_0$ - $\omega_p$. To plot the $p_{in} > p_{crit}$ case, we used arc-length method \cite{Cont_method}}. 
\end{figure}

\noindent JPA enters the bistable regime as the pump power increases larger than the critical value of the pump power ($p_{crit}$) as shown in the figure \ref{fig:1}. For JPA to operate correctly, $p_{in} < p_{crit}$ should be satisfied. We can define the incoming field $a_{in}(t)$ [or intracavity field $A(t)$] as two separate waves which are composed of  pump and signal with the pump coefficient $p_{in}$ [or $p$] being constant and real valued and the signal coefficient $c_{in}(t)$ [or $c(t)$]  being time dependent and complex valued. 
\begin{equation}
\begin{aligned}
a_{\text{in}}(t) = \begin{bmatrix} 
p_{in}e^{-i(\omega_p t + \phi_p)} & c_{\text{in}}(t)e^{-i\omega_p t} 
\end{bmatrix}
\end{aligned}
\label{Freq domain}
\end{equation}

\begin{equation}
\begin{aligned}
A(t) = \begin{bmatrix} 
pe^{-i(\omega_p t + \phi_p)} & c(t)e^{-i\omega_p t} 
\end{bmatrix}
\end{aligned}
\label{Freq domainA}
\end{equation}

\noindent After determining the pump photon number in the steady state, one can solve the QLE equation (\ref{IO theory}) by substituting equations (\ref{Freq domain}) and (\ref{Freq domainA}) in the presence of a weak signal tone $c_{in}$ and we arrive at the equation of motion for the signal field in the time domain:

\begin{equation}
\begin{aligned}
    - i \omega_{p} c(t) + \frac{dc}{dt} =- i \omega_{0} c(t) - i K \left( c^{\dagger}(t) N e^{-2 i \psi_{B}} + 2 c(t) N \right)\\ - \sqrt{2 \gamma_{1}} c_{in}(t) - (\gamma_{1} + \gamma_{2}) c(t)
\end{aligned}
\label{equation of motion time domain}
\end{equation}

\noindent Equation (\ref{equation of motion time domain}) can be solved using numerical methods, or as an integral equation. However, the most pronounced way to solve this equation in the literature is by moving into frequency domain since this is a linear equation and this approach allows one to derive the following analytical expression. 

\begin{equation}
\begin{aligned}
c_{in}(\omega) &= \frac{1}{\sqrt{2\pi}} \int_{-\infty}^{\infty} dt \, c_{in}(t) e^{i\omega t}, \\
c_{out}(\omega) &= \frac{1}{\sqrt{2\pi}} \int_{-\infty}^{\infty} dt \, c_{out}(t) e^{i\omega t}, \\
c(\omega) &= \frac{1}{\sqrt{2\pi}} \int_{-\infty}^{\infty} dt \, c(t) e^{i\omega t}.
\end{aligned}
\label{Fourier definitions}
\end{equation}

\noindent With the Fourier definitions, the equation of motion (\ref{equation of motion time domain}) can be written in the frequency domain as: 

\begin{equation}
\begin{aligned}
    i\left[ (w_{0}-w_{p})-\omega-i(\gamma_{1}+\gamma_{2})+2KN\right] c(\omega)\\+iKNe^{-2i\phi_{B}}c^{\dagger}(-\omega)=-\sqrt{2\gamma_{1}}c_{in}(\omega)
\end{aligned}
\label{EOM Frequency domain}
\end{equation}
\noindent To simplify the notation, the following parameters are defined:

\begin{equation}
\begin{aligned}
    W &= i(\omega_0 - \omega_p) + (\gamma_1 + \gamma_2) + 2iKN \\
    V &= iKN e^{-2i\phi_B}
\end{aligned}
\label{variables}
\end{equation}

\begin{equation}
\begin{aligned}
\lambda_{\pm} = (\gamma_1 + \gamma_2) \pm \sqrt{K^2 N^2 - (\omega_0 - \omega_p + 2KN)^2}
\end{aligned}
\label{lambda}
\end{equation}
\noindent With the definition of the following parameters, when plugged in, we obtain a complex equation. This equation can be solved by considering its hermitian conjugate pair.
\begin{equation}
\begin{aligned}
(W - i \omega ) c ( \omega ) + V c ^ { \dagger } ( - \omega ) = - \sqrt { 2 \gamma _ { 1 } } c_{in} ( \omega )
\end{aligned}
\label{variables}
\end{equation}

\noindent The solution of a quadratic equation yields the lambda value.

\begin{equation}
\begin{aligned}
c(\omega) = \frac{-\sqrt{2\gamma_1} \left[ (W^* - i\omega) c_1^{\text{in}}(\omega) - V c_1^{\text{in} \dagger}(-\omega) \right]}
{(i\omega - \lambda_-) (i\omega - \lambda_+)}.
\end{aligned}
\label{gaineq}
\end{equation}

\noindent This equation describes the evolution of the intracavity field. To compute output field, one may consider boundary condition given in equation (\ref{bound_cond}).

\begin{equation}
\begin{aligned}
c_{\rm out}(t) - c_{\rm in}(t) = \sqrt{2\gamma} c(t)
\end{aligned}
\label{bound_cond}
\end{equation}
\noindent Using boundary condition, the gain equation reads:
\begin{equation}
\begin{split}
c_{\mathrm{out}}(\omega)
= {} & \left(1 - \frac{2\gamma_1(W^* - i\omega)}{(i\omega - \lambda_-)(i\omega - \lambda_+)}\right)
c_{\mathrm{in}}(\omega) \\
&+ \frac{2\gamma_1 V}{(i\omega - \lambda_-)(i\omega - \lambda_+)}
c_{\mathrm{in}}^\dagger(-\omega) \\
\equiv {} & \mathcal{G}(\omega)\,c_{\mathrm{in}}(\omega)
+ \mathcal{M}(\omega)\,c_{\mathrm{in}}^\dagger(-\omega)
\end{split}
\label{finalgaineq}
\end{equation}

 \noindent With equation (\ref{gaineq}), one can calculate the gain. As a result of working in the pump frame, the mode detunings are with respect to the pump frequency. Therefore, \( c_1^{\text{in}}(\omega) \) represents the signal mode and the mode \( c_{in}^{\dagger}(-\omega) \) is the image mode. Defining these modes, the gain is calculated as square of their coefficients.

\begin{equation}
\begin{aligned}
\text{a) } & G_s(\omega) = |{\cal G}(\omega)|^2 \\
\text{b) } & G_i(\omega) = |{\cal M}(\omega)|^2
\end{aligned}
\label{gaindefinition}
\end{equation}

\begin{figure}[H]
\centering
\includegraphics[width=\linewidth,
trim=3cm 9cm 4cm 10cm, clip]{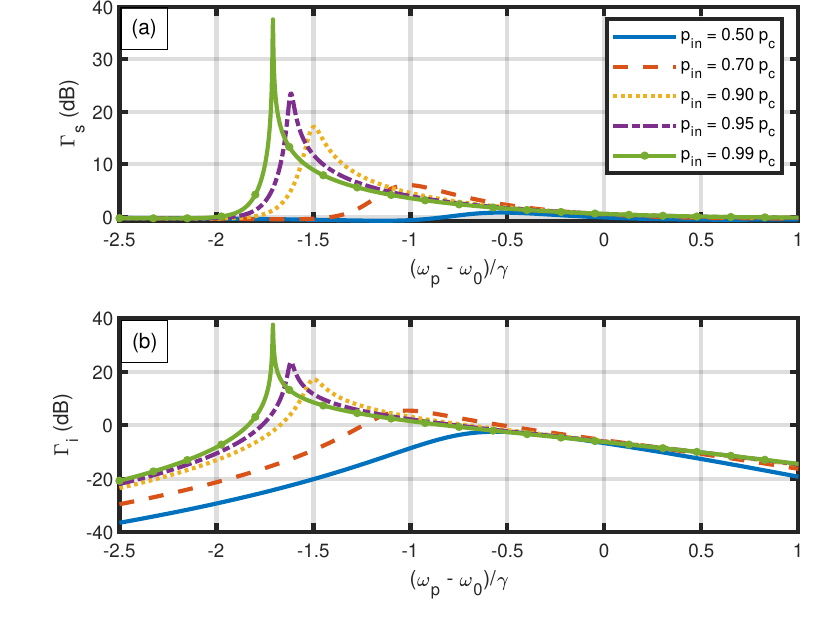}
\caption{\label{fig:3}The solution of Equation (\ref{finalgaineq}) for the degenerate signal mode $(\omega = 0)$ is used to find the optimal pump detuning for a given pump power. (a) Signal gain vs pump detuning, (b) image gain vs detuning
}
\end{figure}

\noindent When one computes optimal pump frequency and power for desired gain from signal gain - pump detuning graph, we can plot signal gain graph spectrum for the device.

\begin{figure}[H]
\centering
\includegraphics[width=\linewidth]{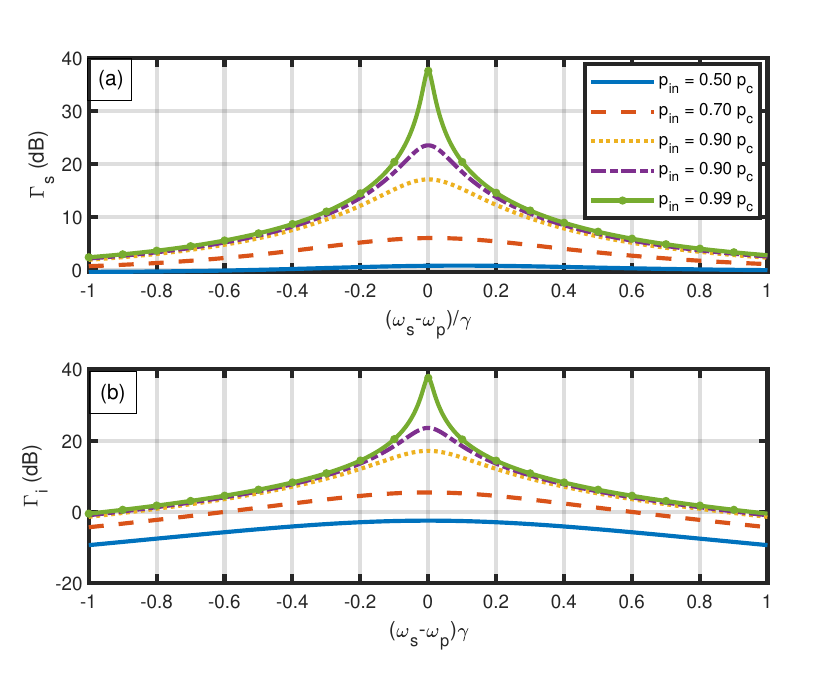}
\caption{\label{QLEW}The solution of Equation (\ref{finalgaineq}) for optimal pump frequency. (a) signal gain vs. signal frequency (optimal $\Delta \omega_p$), (b) image gain vs. signal frequency (optimal $\Delta \omega_p$)
}
\end{figure}

\noindent Figure \ref{QLEW} illustrates both the symmetry and the gain-bandwidth tradeoff. Additionally, quantum theory predicts that a maximum gain of 38dB is theoretically attainable if the biasing is sufficiently precise. However, achieving this regime in practice is challenging because flux or charge noise induced in the JPA can easily push the device into the bifurcation regime.

\section{JPA Design Based on Circuit Model}

\begin{figure}[t]
  \centering
  \begin{subfigure}[b]{\linewidth}
    \includegraphics[width=0.9\linewidth, trim=0.7cm 0.4cm 0.7cm 0.5cm,clip]{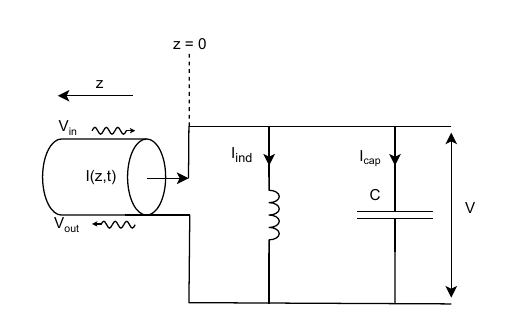}
    \caption{Linear cavity coupled to transmission line.}
    \label{fig:4a}
  \end{subfigure}
  \hfill
  \begin{subfigure}[b]{\linewidth}
    \includegraphics[width=0.9\linewidth, trim=0.8cm 0.4cm 0.7cm 0.5cm,clip]{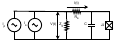}
    \caption{Circuit model used in time-domain analysis, where the Josephson junction replaces the inductor. \cite{Beltran_Tez}}
    \label{fig:4b}
  \end{subfigure}
  \caption{(a) Linear resonator and its coupling to a transmission line. (b) Norton-equivalent model for the JPA used to extract intra-cavity voltage and current.}
  \label{fig:4_combined}
\end{figure}

After calculating the gain of the JPA as explained in Section~\ref{IOT}, in this section we compute it using classical circuit and electromagnetic theory. Since the JPA is a single-port device, it can be described by its one-port reflection coefficient \(S_{11}\). To compute the reflection coefficient, we begin by applying Kirchhoff's Current Law (KCL) to the circuit shown in Fig.~\ref{fig:4_combined}.

\noindent to calculate the reflection coefficient, we write the wave expressions for the transmission line assuming left- and right-traveling waves:
\begin{equation}
V(z,t)=V_{\mathrm{in}}(t - z/v) + V_{\mathrm{out}}(t + z/v),
\label{VoltageDef}
\end{equation}
\begin{equation}
I(z,t)=\frac{1}{Z_c}\left[V_{\mathrm{in}}(t - z/v) - V_{\mathrm{out}}(t + z/v)\right],
\label{CurrentDef}
\end{equation}
where \(v\) is the wave propagation speed and \(Z_c\) is the characteristic impedance. Since the circuit elements are much smaller than the wavelength of operation, we work in the lumped limit and eliminate spatial dependence. The input and output fields are then expressed in terms of the intra-cavity voltage \(V(t)\) and current \(I(t)\):
\begin{align}
V_{\mathrm{in}}(t)&=\frac{V(t)+Z_cI(t)}{2}, \\
V_{\mathrm{out}}(t)&=\frac{V(t)-Z_cI(t)}{2}.
\label{Voltageinandout}
\end{align}

The circuit in Fig.~\ref{fig:4_combined} cannot be directly simulated as a wave system without using an LC ladder model or solving the telegrapher’s equations, both of which are time-consuming and unnecessary in the lumped-element limit. Therefore, we adopt the Norton equivalent circuit (Fig.~\ref{fig:4b}) to obtain intra-cavity quantities. Using Eq.~\eqref{Voltageinandout}, the input and output fields follow, which allows computing the reflection coefficient. Because JoSIM operates in the time domain, we extract specific frequency components via Fourier projection:
\begin{equation}
F(\omega) = \frac{1}{T} \int_{0}^{T} f(t) e^{i \omega_s t} \, dt,
\label{FT}
\end{equation}
which yields the desired frequency-domain signal. The reflection coefficient is defined as the ratio of output to input voltage in the frequency domain \cite{Pozar}:
\begin{equation}
S_{11}(\omega) = \frac{V_{\mathrm{out}}(\omega)}{V_{\mathrm{in}}(\omega)}, \quad 
S_{11}(\mathrm{dB}) = 20 \log_{10} \left| S_{11}(\omega) \right|.
\label{S11definition}
\end{equation}

\begin{figure}[t]
  \centering
  \includegraphics[width=\linewidth, trim=3cm 9cm 4cm 10cm, clip]{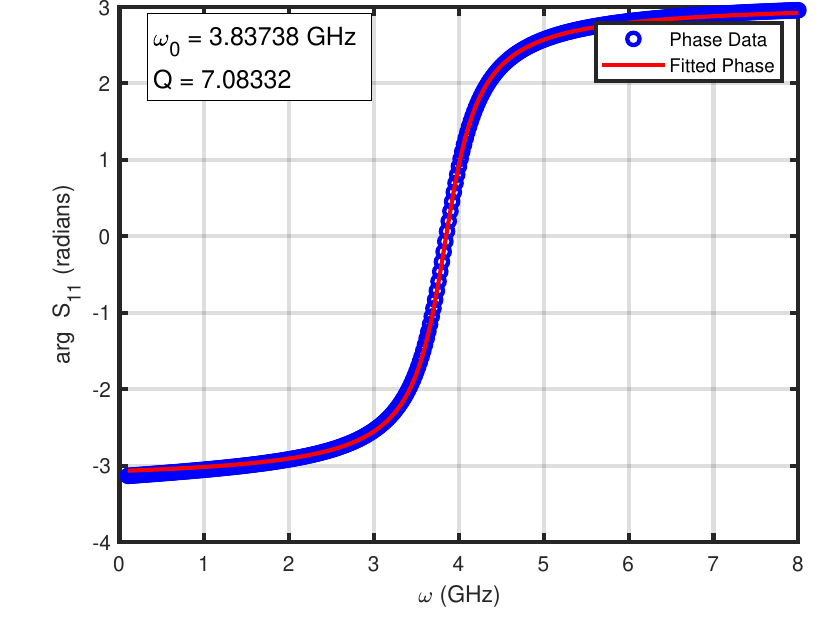}
  \caption{Curve fitting to the reflection phase using Eq.~\eqref{RefCoeff}.}
  \label{fig:2}
\end{figure}

\begin{figure}[t]
  \centering
  \includegraphics[width=\linewidth, trim=3cm 9.5cm 4cm 10cm, clip]{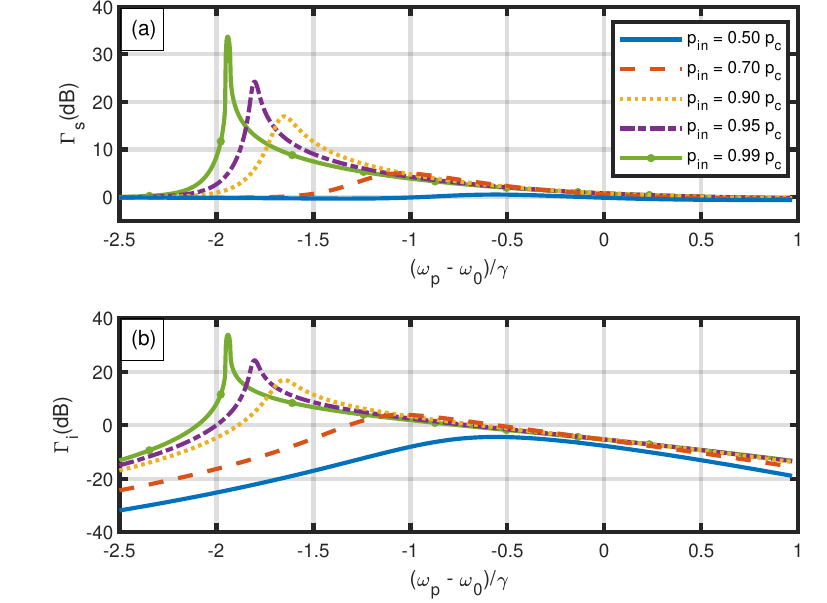}
  \caption{JoSIM gain curve for the degenerate signal mode \((\omega = 0)\), used to find the optimal pump detuning for a given pump power. (a) Signal gain vs.\ pump detuning, (b) idler gain vs.\ pump detuning.}
  \label{fig:JOSIM_wp-w0}
\end{figure}

To characterize the cavity we extract its resonance frequency and quality factor. In the high-\(Q\) limit, the impedance can be approximated as \cite{Beltran_Tez}:
\begin{equation}
Z(\omega) 
= \frac{jL\omega_0^2}{(\omega_0 - \omega)(\omega_0 + \omega)}
\approx \frac{jL\omega_0^2}{2(\omega_0 - \omega)}
= \frac{1}{2C(\omega - \omega_0)},
\label{Impedance}
\end{equation}
where \(L\) is the Josephson inductance, \(\omega_0\) is the resonance frequency, and \(Z(\omega)\) is the resonator impedance. Under weak driving, the JPA behaves approximately linearly. The reflection coefficient of the resonator becomes:
\begin{equation}
\Gamma(\omega) = \frac{Z(\omega) - Z_c}{Z(\omega) + Z_c}
\approx \frac{\gamma - j(\omega - \omega_0)}{\gamma + j(\omega - \omega_0)},
\label{RefCoeff}
\end{equation}
where \(\gamma = \omega_0/(2Q)\) is the linewidth. Figure~\ref{fig:2} shows the phase of \(S_{11}\) as a function of frequency. Fitting this curve yields \(\omega_0\) and \(\gamma\), from which the quality factor is obtained.And to directly compare quantum and classical descriptions, we normalize the detuning \((\omega_p - \omega_0)\) by \(\gamma\), where \(\omega_p\) is the pump frequency. The gain versus normalized detuning for different pump powers is then plotted. Since the JPA is a single-port device, a positive reflection coefficient corresponds to gain. Plotting Eq.~\eqref{S11definition} yields the result in Fig.~\ref{fig:JOSIM_wp-w0}.

\begin{figure}[t]
  \centering
  \includegraphics[width=\linewidth, trim=3cm 9.5cm 4cm 10cm, clip]{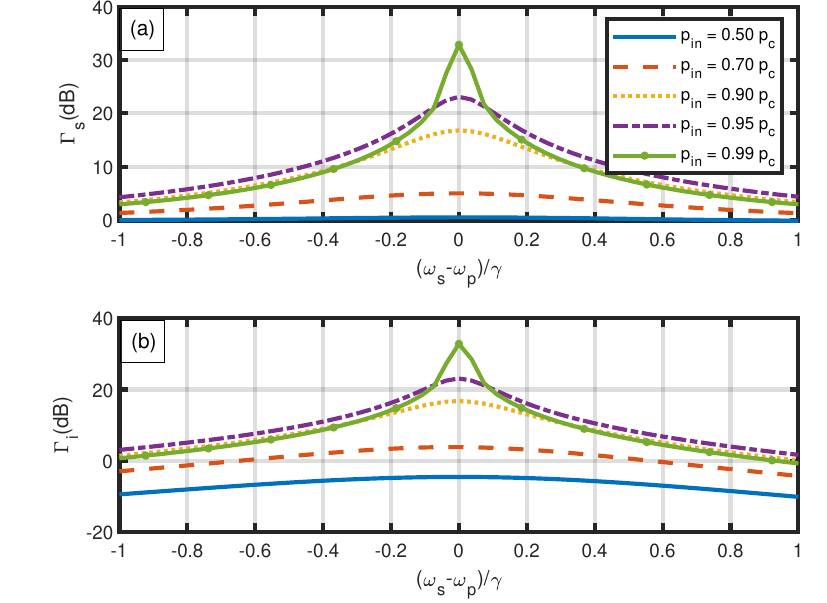}
  \caption{JoSIM gain curve for optimal pump frequency: (a) signal gain vs.\ signal frequency (optimal \(\Delta \omega_p\)), (b) idler gain vs.\ signal frequency (optimal \(\Delta \omega_p\)).}
  \label{fig:JOSIM_ws-w0}
\end{figure}

Choosing \(\omega_p\) at the maximum gain point for each pump power produces the optimal-frequency gain curves shown in Fig.~\ref{fig:JOSIM_ws-w0}.

Figures~\ref{fig:JOSIM_wp-w0} and \ref{fig:JOSIM_ws-w0} show that the gain increases as \(p_{\mathrm{in}}\) approaches \(p_{\mathrm{crit}}\), and the measured gains agree well with the quantum theory apart from slight discrepancies due to the linearization of the quantum Langevin equation and the relatively low quality factor of the cavity \cite{Higher_Order}.

\begin{figure}[t]
  \centering
  \includegraphics[width=\linewidth, trim=3cm 9cm 4cm 10cm, clip]{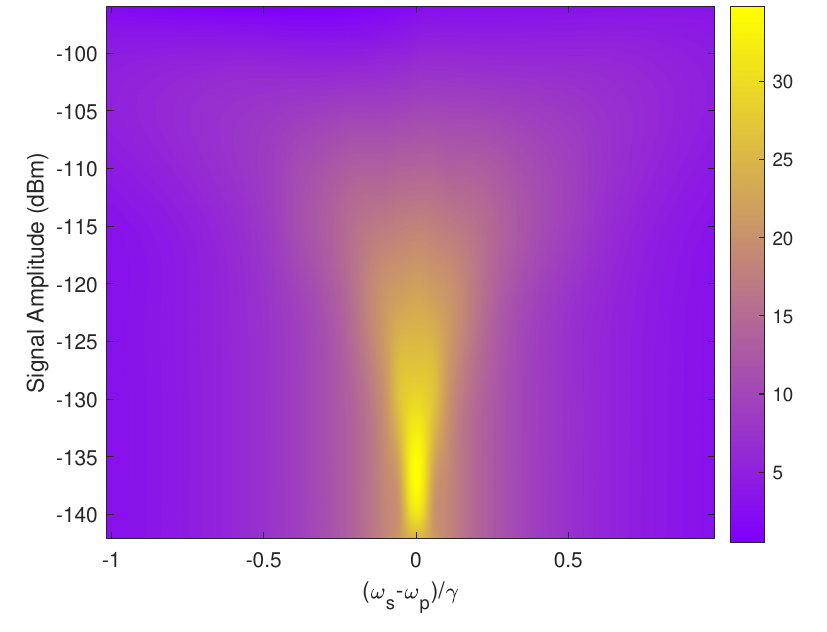}
  \caption{Signal amplitude–signal detuning sweep for pump power \(0.99\,p_c\) and \(\omega_p = 3.305\,\text{GHz}\). The colormap represents signal gain in dB.}
  \label{fig:2Dsweep}
\end{figure}

The saturation power of individual Josephson amplifiers is lower compared to arrays of such devices \cite{Planat_Sat_PWR}. Computing the 1\,dB compression point at the highest gain yields \(-134\,\mathrm{dBm}\), as shown in Fig.~\ref{fig:2Dsweep}. The figure also highlights the gain–bandwidth trade-off in the large input power regime.

\section{Summary and Discussion}

\noindent We computed the gain of the JPA using both quantum and classical frameworks. In the quantum approach, we first derive the system Hamiltonian and model the cavity as weakly coupled to its environment, which enables us to express the intra-cavity field in terms of the input and output fields. This formulation allows for gain calculation either analytically or numerically in the frequency domain. For the classical time-domain analysis, we employ circuit-level models of the JPA components to capture the parametric amplification behavior. Through simple transformations, the gain is extracted from lumped-element simulations. The JPA gain is calculated from the reflection coefficient of the signal wave and its image tone. 

\noindent In summary, we present a systematic way to tackle resonant systems and characterize their reflection coefficient using a circuit simulator, and we show that if modeled correctly, quantum input-output theory and Josephson circuit simulators can be used interchangeably. The use of circuit simulators enables designers to simplify and improve their workflow while advancing the development of complicated quantum circuits.

\section*{Acknowledgment}
The authors thank Nazif Orhun Tekci for his help in editing the figures.

\bibliographystyle{IEEEtran}
\bibliography{JPARefs}

\appendix
\section{Appendix}

\begin{figure}[htbp]
\centering
\includegraphics[width=\linewidth,
trim=0.4cm 0.3cm 0.4cm 0.3cm, clip]{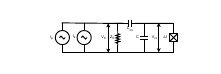}
\caption{\label{MatlabJossimCirc} Josephson circuit model for time domain analysis}
\end{figure}

\noindent In addition to the Josephson circuit based simulators for time domain analysis, we computed the signal characteristics of the JPA starting with the Josephson equations and solve the coupled differential equations by using the ODE45 solver of Matlab with strict tolerences. 
For this purpose, we setup the circuit shown in the figure \ref{MatlabJossimCirc}. The circuit is composed of an input current ($I_{in}$), a pump current ($I_{p}$), a coupling capacitor ($C_{co}$), and a basic JPA structure represented by ($J_{1}$ and $C$). $Z_{c}$ represents the characteristics impedance of the transmission line between the input/pump currents and the JPA. The critical current of $J_{1}$ is $I_{c}$. 

\noindent Starting from the Josephson equations and summing the currents at the input and output nodes of the circuit and solving the equation for the intracavity node flux ($\phi_{int}$), we obtain equation (\ref{eq:third_order})

\begin{equation}
\begin{split}
\frac{d^3\phi_{\text{int}}}{dt^3} ={} & -\frac{C + C_{co}}{C_{co} Z_0 C}\frac{d^2\phi_{\text{int}}}{dt^2}
- \frac{I_c}{C\phi_0} \frac{d\phi_{\text{int}}}{dt} \cos\!\left(\frac{\phi_{\text{int}}}{\phi_0}\right) \\
& + \frac{I_{\text{in}}\omega}{C}\cos(\omega t)
+ \frac{I_p\omega_p}{C}\cos(\omega_p t)
- \frac{I_c}{Z_0 C_{co}} \sin\!\left(\frac{\phi_{\text{int}}}{\phi_0}\right)
\end{split}
\label{eq:third_order}
\end{equation}

\noindent Equation (\ref{eq:third_order}) is the equation of motion of a simple JPA capacitively coupled to a transmission line. Where, $\phi_{0}$ represents the single flux quantum, $\omega_p$ is the pump frequency, and $\omega$ is the signal frequency. By solving the equation, one can compute the transmission line voltages \( V_{\text{tl}} \) and intracavity voltage \( V_{\text{int}} \) as follows:

\begin{equation}
V_{\text{tl}} = -Z_0 C\frac{d^2\phi_{int}}{dt^2} - Z_0 I_c \sin\left(\frac{\phi_{int}}{\phi_0}\right) + (I_{\text{in}} + I_p) Z_0
\label{eq:tlVoltage}
\end{equation}

\begin{equation}
V_{\text{int}} = \frac{d\phi_{int}}{dt}
\label{eq:IntVoltage}
\end{equation}

\begin{figure}[htbp]
\centering
\includegraphics[width=\linewidth]{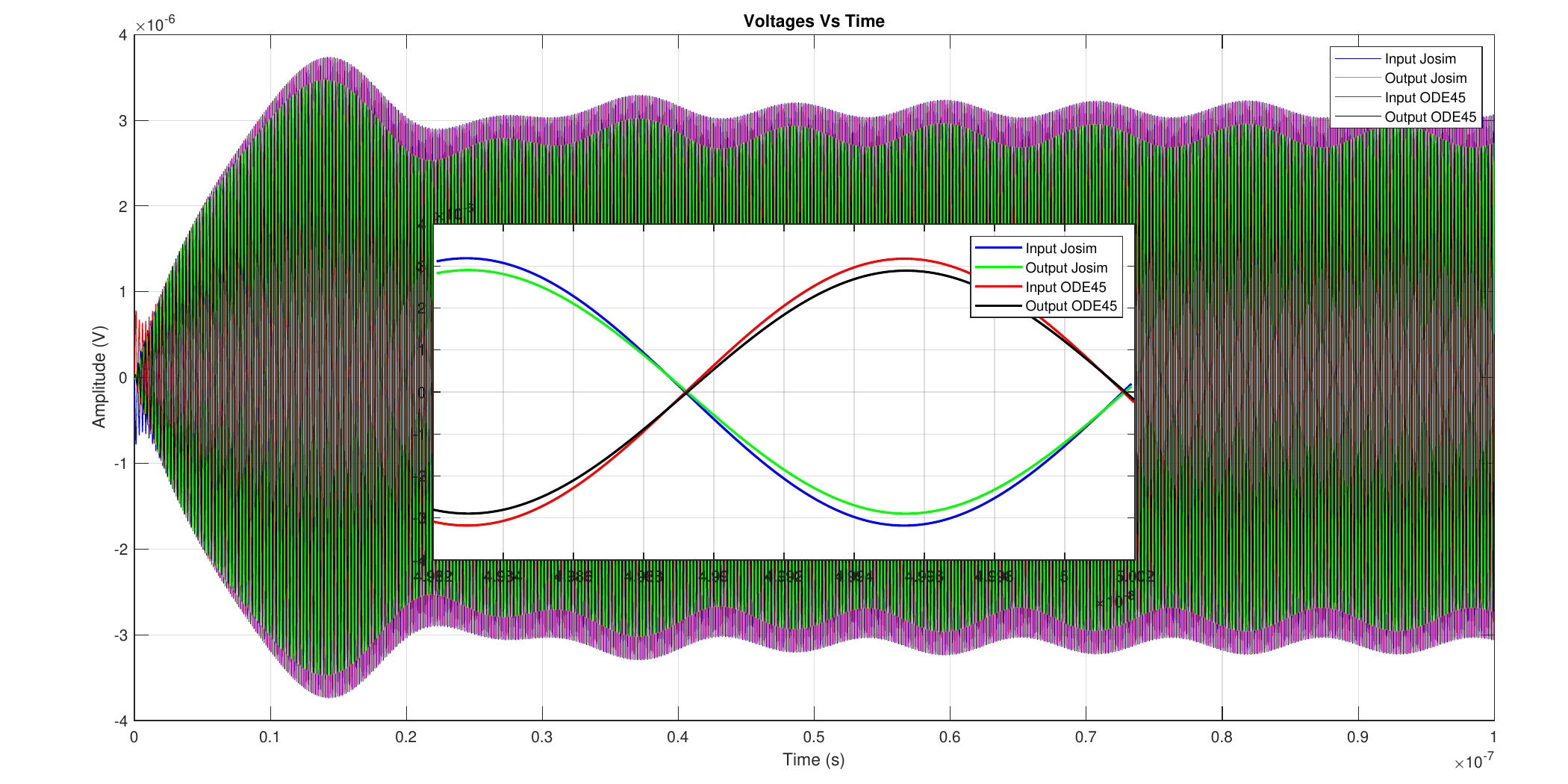}
\caption{\label{MatlabJosim} Comparison between BDF-2 based JoSIM and RK4 Dormand-Prince based ODE45. Inset shows one period of the input and output signals.}
\end{figure}

\noindent In (\ref{eq:tlVoltage}) and (\ref{eq:IntVoltage}), $V_{\text{tl}}$ and $V_{\text{int}}$ also represent the input and output voltages, respectively. If we plot them vs. time, we get the figure \ref{MatlabJosim}. 

\noindent If we compare the discrepancy between two solvers, we get a discrepancy of \%0.36. So, both classical solutions fit with each other and can be used interchangeably where convenient. For the sake of simplicity, we compared the classical solution to quantum solution by using JoSim results. 

\noindent  JoSIM, a time-domain Josephson circuit simulator, employs the BDF-2 method to compute equation derivatives for improved numerical stability. At its core, it solves the standard Kirchhoff equations for capacitors and inductors, with the primary distinction being that calculations are performed in the phase basis. In addition, it incorporates the Josephson relations to accurately model the behavior of Josephson junctions.

\end{document}